# Robust ferromagnetism in insulating $La_2NiMnO_6$ thin films

G. De Luca[1,a)], J. Spring[1], U. Bashir[1], M. Campanini[2], R. Totani[1,b)], C. Dominguez[3], A. Zakharova[4], M. Döbeli[5], T. Greber[1], M. D. Rossell[2], C. Piamonteze[4], M. Gibert[1]

[1] Physik-Institut, University of Zurich, Winterthurerstrasse 190, CH-8057 Zurich, Switzerland.
[2] Electron Microscopy Center, Empa, Swiss Federal Laboratories for Materials Science and Technology, Überlandstrasse 129, CH-8600 Dübendorf, Switzerland.
[3] Department of Quantum Matter Physics, University of Geneva, Quai E.-Ansermet 24, CH-1211 Geneva, Switzerland.
[4] Swiss Light Source, Paul Scherrer Institut, CH-5232 Villigen PSI, Switzerland.
[5] Laboratory of Ion Beam Physics, ETH Zürich, Otto-Stern-Weg 5, CH-8093 Zürich, Switzerland.

*The field of oxide spintronics can strongly benefit from the establishment of robust ferromagnetic insulators with near room-temperature Curie temperature. Here we investigate the structural, electronic, and magnetic properties of atomically-precise epitaxially-strained thin films of the double perovskite $La_2NiMnO_6$ (LNMO) grown by off-axis radio-frequency magnetron sputtering. We find that the films retain both a strong insulating behavior and a bulk-like Curie temperature in the order of 280 K, nearly independently from epitaxial strain conditions. These results suggest a prospective implementation of LNMO films in multi-layer device architectures where a high-temperature ferromagnetic insulating state is a prerequisite.*

[a)] Electronic mail: deluca@physik.uzh.ch
[b)] current address: CNR-IOM, Elettra Sincrotrone Trieste, I-34149 Basovizza, Italy.

# I. Introduction

The development of robust and easy to integrate ferromagnetic insulators (FMIs) is a crucial objective for the progress of next-generation dissipation less oxide spintronic devices [1–3]. Unfortunately, the coexistence of these features is rarely found in nature, mainly because the electrons that mediate FM interactions are typically itinerant [4]. Transition metal oxides, and in particular perovskite-based systems, have been widely studied in the last decades because of the vast combination of functionalities that can be hosted in a relatively simple structure [5,6]. Recent efforts, indeed, have demonstrated that doping, off-stoichiometry, epitaxial strain and interface engineering are examples of feasible strategies to obtain a FMI behavior in perovskite oxide heterostructures [7–10]. These exotic states are, however, in competition with a metallic one suggesting that their adoption in a device architecture could be fragile. In this perspective, it would be more favorable if the nature of the FM interaction was intrinsically insulating.

One possible route to realize this state is provided by double perovskites (DPs). These are compounds characterized by a specific substitution in the $ABO_3$ prototypical perovskite unit cell where exactly half of the A- (B-)site cations are replaced with another cation A' (B'). When these atoms follow a specific long-range order (rock-salt, layered or columnar), the resulting structure is known as A-(B-)site ordered DP and the new compound is characterized by the chemical formula $AA'B_2O_6$ ($A_2BB'O_6$) [11]. A sketch of a B-site rock-salt-ordered DP structure with the pseudocubic cell overlaid in blue is shown in Figure 1a. Due to their structure being similar to standard perovskites, these compounds can be directly integrated into conventional oxide heterostructures and, therefore, investigated using state-of-the-art growth and characterization techniques highly-developed in the perovskite-oxide community.

In standard $ABO_3$ perovskites, the superexchange interaction is often antiferromagnetic (negative) [12]. In B-site ordered DPs, instead, as summarized by the Goodenough, Kanamori and Anderson rules [13,14], the two distinct B and B' cations can be selected such that an empty d orbital of one 3d transition metal interacts with a half-filled 3d orbital of the other cation to produce a positive superexchange interaction. As this particular type of exchange is not mediated by conduction electrons, the resulting ferromagnetic order would naturally emerge within an insulating framework.

Among all the possible choices of A, B and B' cations, a promising configuration is stabilized in $La_2NiMnO_6$ (LNMO). LNMO is an insulator with a relatively high Curie temperature around 280 K [15,16]. It recently attracted further attention due to the discovery of some additional functionalities like colossal magnetodielectricity [17], possible A-site driven ferroelectricity [18], predicted multiferroicity in artificial superlattices [19] or high-temperature paramagnetic spin pumping [20]. In view of the recent demand of FMIs for prospective spintronic applications, it seems timely to reinvestigate the properties of this compound in thin films. Previous reports have shown that both pulsed laser deposition [21–23] and molecular beam epitaxy [24] can be suitable growth methods to produce LNMO thin films. Despite that, to our knowledge, there are no reports systematically investigating the effect of epitaxial strain on the electronic and magnetic properties of stoichiometric LNMO and/or showing the stabilization of a two-dimensional growth mode, a prerequisite for any possible implementation of LNMO thin films in a multi-layer device architecture.

In the following, we will show that atomically precise LNMO films can be deposited by off-axis radio-frequency (RF) magnetron sputtering on a variety of oxide substrates inducing both compressive and tensile strain. We find that their Curie temperature is uncorrelated with epitaxial strain and that the saturation magnetization approaches the bulk value while the films retain insulating transport properties. These results suggest that the magnetic interaction between Ni and Mn in the LNMO system is pretty robust against external perturbations and thus suitable for prospective spintronics devices where a high-temperature FM insulating state is essential.

## II. Structural and electronic properties

LNMO films were epitaxially grown on (001)-oriented $LaAlO_3$ (LAO), $(LaAlO_3)_{0.3}(SrAl_{0.5}Ta_{0.5}O_3)_{0.7}$ (LSAT), $SrTiO_3$ (STO) and (110)-oriented $NdGaO_3$ (NGO), $LaGaO_3$ (LGO) and $DyScO_3$ (DSO) substrates. The films were grown using off-axis RF magnetron sputtering at a substrate temperature of 720 °C and a total pressure of 0.18 mbar in a controlled mixture of oxygen and argon. Additional details on both the growth and characterization methods are discussed in the supplementary material.

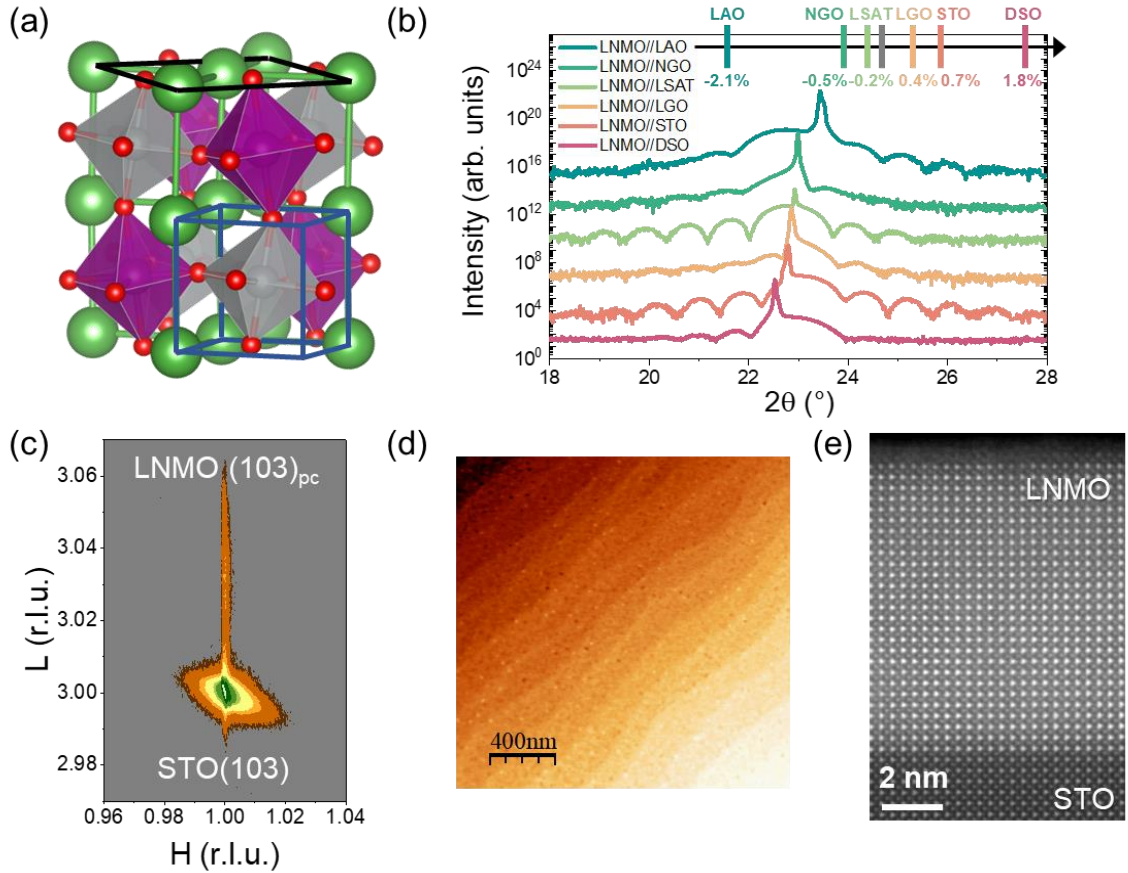

**FIG. 1**. Structural characterization of epitaxially strained LNMO thin films. (a) Schematic of a DP unit cell. The base of the orthorhombic (monoclinic) structure is defined by the black line. The pseudocubic unit cell is shown in blue. (b) X-ray θ-2θ diffractograms of various 30 uc LNMO films around the $(001)_{pc}$ planes of the respective substrates. Fringes are due to finite size effects and have been used to evaluate the film thickness. Different data sets are vertically shifted for better visibility. On top, the nominal strain exerted on the epitaxial LNMO films by the distinct substrates is highlighted. (c) Reciprocal space map around the $(103)_{STO}$ diffraction condition certifies the fully-strained state of the LNMO film. (d) Surface topography of a representative LNMO//STO film as obtained by AFM. (e) STEM image along the $[100]_{STO}$ zone axis indicates a high crystalline quality of the film and the absence of parasitic defects.

Assuming a pseudocubic lattice constant of LNMO of 3.876 Å (a value obtained averaging the characteristic lattice constant of the two room-temperature LNMO polymorphs, rhombohedral and monoclinic [25]), a compressive strain of -2.1%, -0.5%, -0.2% on LAO, NGO, LSAT and a tensile strain of +0.4%, +0.7% and +1.8% on LGO, STO and DSO substrates, respectively, is imposed. X-ray diffraction (XRD) θ-2θ scans around the pseudocubic (pc) (001) diffraction plane of the aforementioned substrates is shown in Figure 1b and highlights the crystalline quality of our sputter-grown films. From the fringes in the XRD data we confirmed that the thickness of our strained films is in the order of 30 pc unit cells (uc), ≈ 12 nm, as expected from previous growth calibrations. All six films are fully-strained to the substrate lattice constant as observed in the reciprocal space maps measured around the $(103)_{pc}$ diffraction condition of the respective substrates (Figure 1c for LNMO//STO and Figure S1 for a selection of other strained films).

In Figure 1d we present the surface topography of a representative LNMO//STO film imaged by atomic force microscopy (AFM). Atomic steps inherited from the underlying substrate topography suggest the stabilization of a two-dimensional growth mode. Similar step-and-terrace morphology is also obtained for films grown on LSAT, NGO, LGO and DSO while the topography of a LNMO//LAO film is characterized by island growth (not shown). This can be tentatively attributed to the large compressive strain exerted by the substrate that might require substrate-specific growth optimizations [26]. Finally, in Figure 1e, we show a scanning transmission electron microscopy (STEM) image of a LNMO//STO film that highlights the absence of both dislocations and undesired parasitic phases. The stoichiometry of our films was verified using both Rutherford backscattering (RBS) and X-ray photoemission spectroscopy (XPS), as discussed in more detail in the supplementary material.

The evidence of correct stoichiometry, however, is not enough to determine the presence of cation ordering. To validate that our LNMO films are not a solid solution but rather formed by a rock-salt arrangement of the $NiO_6$ and $MnO_6$ octahedra we measured the XRD along the $[111]_{pc}$ direction. In Figure 2a we show the scan performed along this direction for a 35 nm (around 90 uc) LNMO film grown on a (001)-oriented STO substrate. The same measurement for a bare STO substrate is also displayed for comparison. As a direct consequence of the rock-salt ordering of the B and B' cations forming the DP cell, a new periodicity along $(½, ½, ½)_{pc}$ emerges in our film, which confirms the long-range order of Mn and Ni cations [27]. With the aid of selection rules [28], we can exclude that this diffraction peak emerges from the octahedra tilt system of LNMO. It has been observed, however, that the origin of these superstructures in DPs could also occur due to unequal A-cation displacement along the $[111]_{pc}$ direction, mimicking the unit cell of the chemically-ordered compound [18,29]. As the chemical ordering in LNMO bulk specimens is accompanied by an electronic charge transfer from the nominally $Ni^{3+}$ ions to the nominally $Mn^{3+}$ ions culminating in a $Ni^{2+}/Mn^{4+}$ electronic configuration [17,30], the observation of this charge transfer can also be considered a landmark of the rock-salt ordering of the $NiO_6$ and $MnO_6$ octahedra forming the DP structure [15].

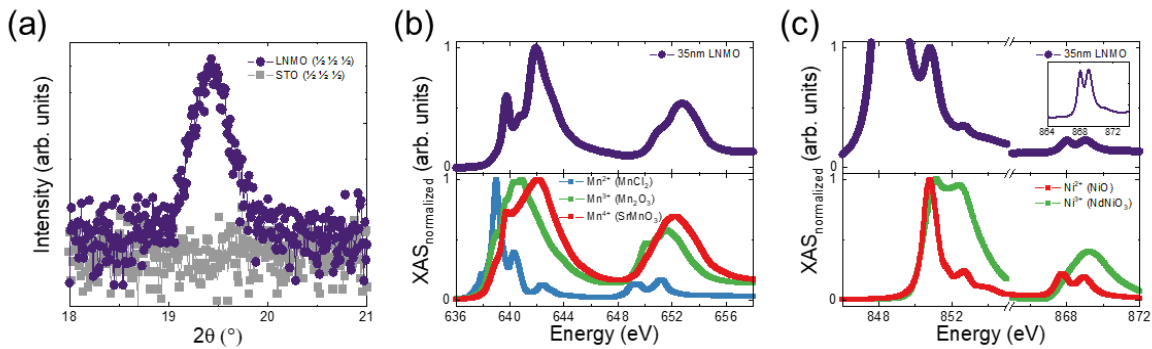

**FIG. 2**. B-cation ordering. (a) XRD collected along the $(111)_{pc}$ plane of a 35 nm LNMO//STO film (purple circles) and a representative STO substrate (grey squares). The half-order peak emerges due to the three-dimensional B-cation-ordering of the DP structure. (b) Top: XAS data collected at 20 K in TEY mode at the Mn $L_{2,3}$-edges of the same 35nm LNMO//STO film normalized to the $L_3$-edge maximum. Bottom: Reference Mn spectra. (c) Same as (b) collected at the Ni $L_{2,3}$-edges. The out-of-scale peak around 849 eV corresponds to the La $M_4$ edge. Comparison between top and bottom panels certifies the prevalence of both $Mn^{4+}$ and $Ni^{2+}$ valence configurations.

To access the electronic configuration of Ni and Mn, we employed X-ray absorption spectroscopy (XAS) in the soft x-ray regime, as this technique is strongly sensitive to the valence state of the probed transition-metal ion [31]. In the top panels of Figure 2b (c) we present the XAS collected at the Mn (Ni) $L_{2,3}$ edges of the same 35 nm-thick LNMO//STO film while, in the bottom panels, we show the associated reference spectra. A comparison between the LNMO spectra and the references indicates that the Mn adopts a 4+ configuration while the Ni is in a 2+ state. This is particularly evident from two specific features: first, the position of the Mn $L_3$-edge pre-peak corresponds to the one of $SrMnO_3$ ($Mn^{4+}$); second, the multiplet structure of the Ni $L_2$-edge is reminiscent of the one measured in NiO ($Ni^{2+}$). The differences between the LNMO and the reference spectra are attributed to the distinct mutual crystal field environment of the Ni and Mn cations, probably inducing a different line broadening [32]. Similar room-temperature XAS spectra are observed for all the 30 uc films grown on different substrates (Figure S2a,b). Hence, we validate that LNMO films grown by off-axis RF magnetron sputtering are stoichiometric and with a cation-ordered DP structure. Finally, we note that the room-temperature resistivity of all investigated LNMO films averages to 900 Ω·cm, certifying a strong insulating behavior [16], independent of the specific strain state.

### III. Magnetic properties

After having determined the structural and electronic properties of our LNMO films, we now investigate their magnetic properties.

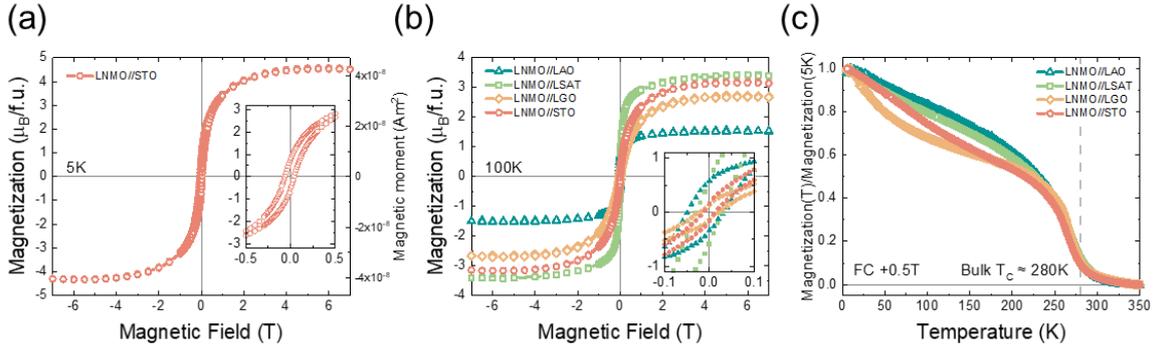

**FIG. 3.** Magnetic properties investigated by SQUID magnetometry. (a) Isothermal magnetization curve collected at 5 K for a 30 uc LNMO//STO film. We observe a saturation magnetization of 4.3 $\mu_B$/formula unit (magnetic moment of $4.2\times10^{-8}$ $Am^2$). Inset: closer view of the opening of the hysteresis loop. (b) Magnetization curve collected at 100 K for films of 30 uc grown on LAO, LSAT, LGO and STO. The film grown on LAO, subjected to a larger epitaxial strain, is characterized by a lower in-plane saturation magnetization. The inset shows that, together with the hysteresis opening, we also observe a slightly larger coercive field for increasing compressive epitaxial strain. (c) The magnetization as a function of the temperature normalized to the 5 K value indicates similar Curie temperatures for different LNMO films, independent from their epitaxial strain. The bulk $T_C$ is highlighted with a dashed line. The data is collected during field cooling (FC) in an in-plane magnetic field of +0.5 T.

We present in Figure 3a, the hysteresis loop of a 30 uc LNMO//STO film measured at 5 K by superconducting quantum interference device (SQUID) magnetometry. Special care was taken to remove the substrate contribution [33]. The film is characterized by a saturation magnetization of 4.3 $\mu_B$/formula unit (f.u.), slightly lower than the originally reported value of 4.9 $\mu_B$/f.u. (3.0 $\mu_B$ on the Mn site and 1.9 $\mu_B$ on the Ni site) obtained after Rietveld refinement of neutron diffraction patterns [16]. We can attribute this small discrepancy to a combination of various effects such as different Mn-O-Ni canting-angle in the strained film [34,35] but also to the presence of residual antisite disorder [15,36]. Nevertheless, similar saturation magnetization values have also been reported in other high-quality bulk specimens [34,36]. We also observe a coercive field in the order of 300 Oe and a low remnant magnetization (< 1 $\mu_B$/f.u), again in accordance with values reported in the literature [15,16]. The low remanence, in particular, has been previously attributed to the presence of long-range ordered FM domains that, in stoichiometric films, are coupled antiferromagnetically across antiphase boundaries [15], a characteristic that LNMO share with other magnetically-ordered DPs [37]. A higher remnant magnetization ($\gtrsim$ 1 $\mu_B$/f.u) can, instead, be associated with an excess of Mn [15], and it should be considered a warning for non-stoichiometric films [21,27,38].

In Figure 3b we compare the hysteresis loops collected at 100 K for some of the differently strained samples. Films grown on LSAT, LGO and STO seem to have similar saturation magnetization, around 3 $\mu_B$/f.u. at 100 K whereas the film grown on LAO has a lower saturation magnetization. The strong paramagnetic signal of both NGO and DSO substrates excludes the possibility of using SQUID magnetometry to characterize the magnetic properties of the associated LNMO films. We additionally observe a small trend of increasing coercive field as compressive strain increases (inset of Figure 3b). This can preliminarily be associated to a hardening of the in-plane magnetic axis induced by epitaxial strain [39,40]. Finally, in Figure 3c, we compare the temperature-dependent magnetization normalized to its 5 K value for the films shown in Figure 3b. We find that, despite the different saturation magnetizations, the LNMO films are always characterized by a Curie temperature around 280 K accompanied by a non-zero magnetization tail at higher temperatures [41–43] to which we will come back shortly. This remarkable stability against epitaxial strain corroborates a similar result obtained in bulk specimens using hydrostatic pressure [44] and further suggests a prospective use of LNMO films for spintronic applications.

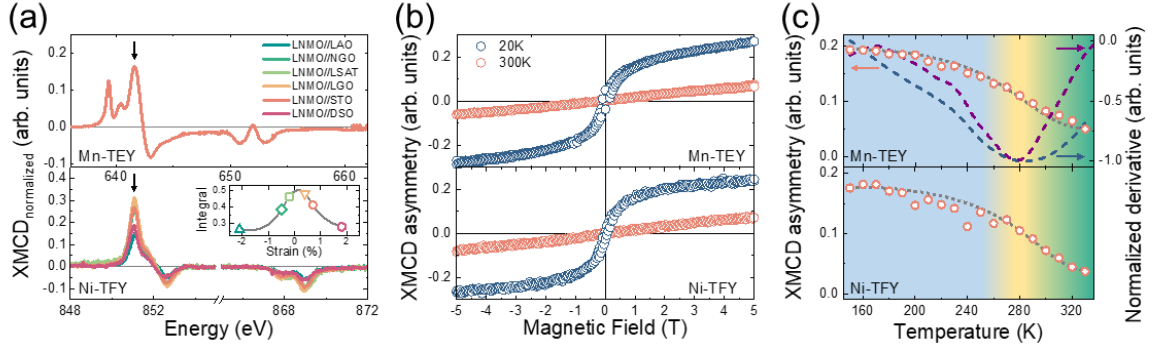

**FIG. 4.** Magnetic properties investigated by synchrotron radiation. (a) In the top (bottom) panel it is displayed the XMCD data collected at the Mn (Ni) $L_{2,3}$-edges at 100 K for a LNMO//STO film in TEY (TFY) mode. For the Ni $L_{2,3}$-edges all the other strained films are also shown (see main text). Black arrows indicate the resonant energy chosen to calculate the asymmetry ratio. In inset, the integral calculated over the Ni $L_{2,3}$-edges is shown, with the grey line being a guide to the eyes. (b) In the top (bottom) panel XMCD asymmetry ratio of Mn (Ni) as a function of the magnetic field measured at 20 K (blue circles) and 300 K (red circles) for the LNMO//STO film. (c) In the top (bottom) panel XMCD asymmetry ratio of Mn (Ni) as a function of the sample temperature measured in a magnetic field of 5 T (red circles). The grey dotted line is a guide to the eyes. Dashed lines indicate the derivative of XMCD asymmetry (blue) and SQUID (purple) data measured with the same magnetic field. The transition interval (yellow) from the ferromagnetic (light blue) to the paramagnetic phase (green) is chosen around the minimum of the derivatives. Colored arrows indicate the respective axes.

To further characterize the magnetism of the strained films we performed X-ray magnetic circular dichroism (XMCD) measurements. Due to the insulating nature of both films and substrates, we observe a strong charging in the data collected in total electron yield (TEY) mode which is promoted by the synchrotron radiation and further increases when lowering the temperature. Only the films grown on STO seem to be less affected by this issue, probably due to the innate ability of this substrate to transport charge carriers [45]. At the Ni $L_{2,3}$-edge, this obstacle can be overcome by measuring in total fluorescence yield (TFY). Unfortunately, this is not the case for the Mn edges due to the dominant role of self-absorption [46] caused by the proximity to the O K-edge [47]. Consequently, we show in Figure 4a the dichroism measured at 100 K and 5 T at the Mn $L_{2,3}$-edges in TEY mode for the LNMO//STO sample (top panel) and at the Ni $L_{2,3}$-edges in TFY mode for all strained films (bottom panel). From the concordant sign of the XMCD measured at both edges, we confirm the expected ferromagnetic coupling of the Ni and Mn sublattices [48]. To qualitatively determine a trend as a function of strain, the inset of Figure 4a displays the integral calculated over the Ni $L_3$-edge subtracted by two times the integral calculated over the Ni $L_2$-edge (the numerator in the spin sum rule [49]). From this, we derive that the saturation magnetization of the LNMO films decreases as either compressive or tensile strain increases, corroborating our previous observation by SQUID magnetometry (see for instance the LNMO//LAO sample in Figure 3b). It is known that perovskites accommodate epitaxial strain also with vacancies formation [50,51]. It is plausible, therefore, that such reduced structural stability could result in an increased antisite disorder leading to additional antiferromagnetic interactions via the $Mn^{4+}$-$O^{2-}$-$Mn^{4+}$ and $Ni^{2+}$-$O^{2-}$-$Ni^{2+}$ bonds [15,36]. Such a lowering of the saturation magnetization, however, seems to be not correlated with a reduction of the Curie temperature [36], as shown in Figure 3c, again remarking the impressive stability of the $Mn^{4+}$-$O^{2-}$-$Ni^{2+}$ ferromagnetic interaction [44,52].

With XMCD, we can also follow the element-specific evolution of the magnetization as a function of the external magnetic field. In particular, we plot in Figure 4b the XMCD asymmetry ratio defined as $\frac{L-R}{L+R}$ and measured at 20 K and 300 K for a 30 uc LNMO//STO film. Here, $L$ ($R$) is obtained as the spectral difference measured at the $L_3$-resonant (indicated by the small arrows in Figure 4a) and off-resonant energies for left (right) circular polarization. The magnetic response of the Mn and Ni sublattices is similar, confirming that despite the DP unit cell being formed by two distinct transition metal ions, it behaves as if featuring a single magnetic sublattice. Moreover, while the asymmetry ratio measured at 20 K tends to saturate, a clear linear paramagnetic shape is observed at 300 K. This confirms that the non-zero room-temperature magnetic moment measured by SQUID (Figure 3c) is not due to ferromagnetic ordering but attributed to a combination of field effect and short-range magnetic correlations [41,43,53].

Finally, to present a complementary way to determine the Curie temperature, we show in Figure 4c the XMCD asymmetry ratio of the same LNMO//STO sample collected with a field of 5 T as a function of the temperature. We use a strong external field to saturate the magnetization and to maximize the signal-to-noise ratio but, as a consequence, the ferromagnetic-paramagnetic phase transition is smeared out. To determine the transition temperature, we calculate the derivative of the XMCD asymmetry and we fix $T_C$ to be its minimum. The Curie temperature derived in this way is located slightly above 280 K. Using the same derivative approach to analyze the SQUID magnetometry data [54] collected in a field of 5 T, we find a $T_C$ around 275 K. Both these numbers are in agreement with the reported bulk values [15,16] and thus corroborate the use of temperature-dependent XMCD to investigate magnetic phase transitions [55,56].

### IV.  Summary and conclusions

We have shown that off-axis RF magnetron sputtering is a suitable growth method to produce stoichiometric and cation-ordered LNMO thin films. We find that our films are characterized by a robust insulating behavior paired with a Curie temperature in the order of 280 K, nearly independent from epitaxial strain. We further observe that films subjected to lower epitaxial strain possess a larger saturation magnetization, approaching the theoretical value of 5 $\mu_B$/f.u. XMCD hysteresis loops and thermal asymmetry collected at both the Mn and Ni edges further clarify the paramagnetic nature of the LNMO films at room temperature. The non-zero magnetization observed at 300 K must be attributed to the presence of short-range correlations [41]. These can still be exploited in potential applications where interfacial effects are dominant [20]. Such a remarkable stability of the ferromagnetic interaction between $Ni^{2+}$ and $Mn^{4+}$ can be highly beneficial for the implementation of LNMO thin films in prospective spintronic devices.

**Acknowledgements**

G.D.L., J.S., U.B. and M.G. thank Scott Chambers for valuable discussions. This research was supported by the Swiss National Science Foundation (under Project No. PP00P2_170564 and Grant Nos. 206021_150784 Requip ASKUZI).